# Chaotic Turing Patterns displaying the beauty of chaos

Jinghua Xiao [1,2], Junzhong Yang [2], Gang Hu [3,1*]

[1] *Department of Physics, Beijing Normal University, Beijing 100875, China*

[2] *School of Science, Beijing University of Posts and Telecommunications, Beijing 100876, China*

[3] *China Center for Advanced Science and Technology (CCAST) (World Laboratory), P. O. Box 8730, Beijing 100080, China*

**Abstract:**

The problem of Turing pattern formation has attracted much attention in nonlinear science as well as physics, chemistry and biology. So far all Turing patterns have been observed in stationary and oscillatory media only. In this letter we find for the first time that ordered Turing patterns exist in chaotic extended systems. And chaotic Turing patterns are strikingly rich and surprisingly beautiful with their space structures. These findings are in sharp contrast with the intuition of pseudo-randomness of chaos. The richness and beauty of the chaotic Turing patterns are attributed to a large variety of symmetry properties realized by various types of self-organizations of partial chaos synchronizations. Some statistical measurements are performed to confirm and heuristically understand these findings.

It has been known for a half of century since the pioneering work of Turing [1] that spatially ordered and localized structures, namely Turing patterns, can appear from stationary homogeneous media by self-organizations and symmetry breakings [1-7]. Turing pattern formation has attracted great attention in nonlinear science as well as physics [2-6], chemistry [7-8] and biology [9-11]. A very recent and significant advance in this direction is that oscillatory Turing patterns were observed at the Turing instabilities of periodic homogeneous media [12-16]. On the other hand, in the past several decades, the issue of chaos in spatiotemporal systems has become one of the focuses in nonlinear science [17-20]. It seems very unlikely that chaotic systems could generate spatially ordered Turing patterns, because temporally chaotic motions (trajectories are sensitive to extremely small perturbations) are not supposed compatible with ordered space structures (which should be stable against rather strong noise). Although some symmetry properties of a few coupled chaotic oscillators have been explored, based on the concept of partial synchronization [21-25], the topic of Turing pattern formation in chaotic extended has not been investigated.

In this work, we find for the first time that spatially ordered Turing patterns exist in chaotic extended systems. In addition, chaotic Turing patterns are strikingly rich and surprisingly beautiful with their space structures. The richness and beauty of the chaotic Turing patterns are due to a large variety of symmetry properties realized by various types of self-organizations of partial chaos synchronizations in the pattern formation.

In this paper, we study the diffusively coupled Rossler system in two-dimensional ($L \times L$) plane with periodic boundary condition and chaotic local dynamics

$$\partial_t u(x,y) = -(v+w) + D_u \nabla u$$
$$\partial_t v(x,y) = u + Av + D_v \nabla v$$
$$\partial_t w(x,y) = uw - Cw + B + D_w \nabla w \quad (1)$$
$$\nabla = \frac{\partial}{\partial x^2} + \frac{\partial}{\partial y^2}$$

Let us start our investigation from chaos synchronization. In Fig. 1(a) we show a critical curve above which the extended system is completely synchronous, i.e., all space points of the medium follow an identical chaotic temporal orbit. Below the threshold synchronous chaos becomes unstable, and various desynchronous and chaotic motions can be observed. In Fig. 1(b) we choose a parameter set below the threshold curve, the space-time waves look chaotic in both time and space. From Fig. 1(b) one can hardly imagine that the chaotic randomness can be associated with any finely ordered spatial structure.

However, the spatial distributions of variables provide completely different pictures. In Figs. 2(a)-(c) we plot the snapshots of spatial distributions of variable $u(x,y,t)$ at different time instants with $40 \times 40$ pixels. The spatial contours show nicely ordered structures in sharp contrast with the chaotic waves of Fig. 1(b). It is clear from these contour figures that after desynchronization of homogeneous chaos the chaotic motions of different space points retain certain partial synchronizations, which provide some well organized spatial structures. There are a number of ordered patterns appearing in the time evolution. As time goes on, the system state varies from one pattern to another chaotically. And it is the chaoticity in pattern alternations that leads to the random-like trajectories of Fig. 1(c). In Fig. 2(a)-2(c) one can observe a characteristic wave length. When we enlarge the system size to $80 \times 80$ pixels with coupling unchanged, we find in Figs. 2(d)-2(f) the same wave length with the number of waves doubled.

The most important characteristic feature of Turing patterns is localization of space variable distributions. In order to test the existence of localized structures of the chaotic patterns we compute a distribution of $u_M(x,y)$ which is the maximum of $u(x,y,t)$ at the space point $(x, y)$ over a long time interval $T$. And a number of distributions of $u_M(x,y)$ are plotted in Fig.3 for different sets of coupling parameters. Unlike the snapshots in Fig.2 showing explicit time variation of $u(x, y, t)$ patterns, the patterns of $u_M(x,y)$ in Fig.3 manifest the asymptotic distributions implicit in the long time statistics (note, further increase of length $T$ does no longer change the contour pictures). Therefore, the figures in Fig.3 show for the first time the localized Turing spots in chaotic media, which are called *chaotic Turing patterns*.

By varying coupling parameters, one can observe a large number of characteristically different chaotic Turing pictures. Some patterns of Fig.3 have not been observed in the conventional Turing patterns of stationary and periodic media. The conventional Turing patterns are determined strictly by one or few modes at instabilities, and these unstable modes produce some typical types of Turing patterns, e.g., hexagonal, squares, rolls, honeycombs, stripes, and their combinative and periodically alternative structures [12-13]. In contrast, in chaotic spatiotemporal systems, owing chaoticity, many modes can be easily coupled and involved in the pattern formation process, and this provides much larger freedom to generate a large variety of distinctive patterns.

In Fig.3 most of pictures have well organized structures rooted by different types of symmetry breakings. The synchronous state of Eq.(1) has full space symmetric properties, and thus does not have any spatial localization structure. The homogeneous distribution is invariant against all the space translation, mirror reflection, and rotation transformation groups. After the instability of the synchronous chaos most of these symmetries are broken, and the symmetries retained in the desynchronous chaos determine the spatial orderings of the inhomogeneous states and result in diverse structures.

In Figs. 3(a) and 3(b), the window-like and flower-like patterns are invariant against the mirror reflections with both *x*- and *y*- axes $[u_M(x,y) = u_M(x,-y) = u_M(-x,y) = u_M(-x,-y)]$. The spontaneous emergences of a symmetry center [set to (*x*, *y*)=(0,0) now] and two symmetry axes are the typical characteristics of these figures. Pattern Fig. 3(c) does not possess any mirror symmetry, but it has discrete translational symmetries along both *x* and *y* directions. And these symmetries yield a carpet-like pattern. The symmetry properties of Fig. 3(d) are different. It has the translational symmetries of $u_M(x+\frac{nL}{2}, y+\frac{nL}{2}) = u_M(x+\frac{mL}{2}, y-\frac{mL}{2}) = u_M(x,y), n,m = \pm 1$ and the reflection symmetry against the symmetry center (0, 0), while it does not possess mirror symmetry against any axis. Figure 3(e) has mirror symmetries against four axis (*x*-, *y*-, *x+y*=0 and *x-y*=0 axes) and the rotational symmetries with $\pi$ and $\frac{\pi}{2}$ angles around the rotation center (0, 0). (Note $\frac{\pi}{2}$ rotational symmetry does not exist in all the previous patterns). The pattern of Fig. 3(f) is similar most to the conventional Turing patterns. It possesses all translational symmetries $u_M(x+\frac{nL}{4}, y+\frac{mL}{4}) = u_M(x,y), n,m = 0,1,2,3$, mirror reflection symmetries (with 8 horizontal and 8 vertical symmetry axes and 64 symmetry centers), and rotational symmetries of π/2 and π rotation angles around all the 64 symmetry centers. We would like to emphasize that a large variety of patterns with different symmetry properties can be observed in a single system of Eq.(1) among which only very few are shown in Fig.3. This richness is in sharp contrast with conventional stationary and periodic systems where a single system usually shows only few characteristically different Turing patterns.

Three points are crucial for characterizing chaotic Turing patters: (i) Chaoticity of temporal motions; (ii) Chaos synchronizations among partial space points; (iii) Space orderings organized by these chaos synchronizations. In Fig.4 we perform some standard numerical and statistical measurements to demonstrate these aspects. We show chaotic orbit in Fig. 4(a) and divergence of adjacent trajectories in 4(b), indicating chaos without ambiguity. In Fig.4(c) we compute a modified correlation function [26] for different space points

$$S(x,y) = \sqrt{\frac{\left\langle \left[u_{x_0,y_0}(t) - u_{x,y}(t)\right]^2 \right\rangle}{\left[\left\langle u^2_{x_0,y_0}(t)\right\rangle \left\langle u^2_{x,y}(t)\right\rangle\right]^{1/2}}} \quad (2)$$

$$\left\langle f_{x,y} \right\rangle = \lim_{T\to\infty} \frac{1}{T} \int_0^T f_{x,y}(t)dt$$

Function (2) is equivalent to standard space correlations, however, the form of Eq.(2)

is particularly useful for quantitatively measuring the degree of chaos synchronization, where we observe $S(x,y) \approx 0$ in a number of space areas, identifying chaos synchronizations. A striking observation is that these synchronous space regions are distantly away from each other and they are desynchronous from all the space sites around them, a typical feature of nonlocal chaos synchronization [27]. In Fig. 4(d) we plot the average absolute differences of space variables

$$\Delta(x,y) = \lim_{T \to \infty} \frac{1}{T} \int_0^T |u(x,y,t) - u(x_0,y_0,t)| dt \qquad (3)$$

In Figs. 4(c) and (d) spatial orderings induced by perfect self-organization of chaos synchronization become apparent. These orderings are irrelevant to the position of the reference point ($x_0, y_0$). Note Figs. (1c), 4(c) and 4(d) represent the same spatiotemporal state. After a long time average, we explore strikingly well ordered space structure of Figs. 4(c) (d) from the pseudo-random chaotic state Fig.1(c).

In conclusion, we have revealed, for the first time, localized structures of Turing patterns in chaotic extended systems. The richness and complexity of the chaotic Turing patterns are due to the diverse symmetry properties realized by partial chaos synchronizations. The present work focuses on the numerical demonstrations of Turing patterns in a single chaotic Rossler system with a specific coupling structure. We find similar phenomena in other systems, such as Rossler systems with different coupling structures and coupled Lorenz equations and Chua's circuit systems and so on. Further explorations of chaotic Turing patterns in natural pattern formation processes and in experimental realizations will stimulate a new area of pattern formation.

This work was supported by the National Natural Science Foundation of China under Grant No. 10335010 and 10172020 and by Nonlinear Science Project.

**Captions of figures**

Fig. 1

Results of numerical simulations of Eq.(1) with A=0.2, B=0.2, C=4.5. 40×40 pixels are applied for discretization of 10×10 physical space. These parameters are used for all the figures unless specified otherwise. In simulations we start from the initial conditions $u(x, y, 0)=v(x, y, 0)=w(x, y, 0)=\delta(x, y)$ with $\delta(x, y)$ being randomly chosen in [0.0, 0.2]. (a) Critical curve in the diffusion parameter plane ($D_u = D_v$) for the stability of synchronous chaos of Eq.(1). Chaotic Turing patterns can be observed below the threshold. (b) $D_u = D_v$=0.005, $D_w$=2.5 (below the critical curve). Chaotic waves in $(t, x)$ plane for y=4.

Fig. 2

Snapshots of contour pictures of the time evolution of $u(x, y, t)$ at different time instants. (a)-(c) Numerical results in $40 \times 40$ pixels (a) $t$ =2000.5 ; (b) $t$ =2001.8; (c) $t$ =2002.1. $D_u, D_v$ and $D_w$ are the same as Figs.1(c). (d)-(f) The same as (a) to (c) with $80 \times 80$ pixels in $20 \times 20$ physical plane used. (a) t=1999.5;(b) t=2000.8;(f) t=2001.1.

Fig.3

Beauty of chaotic Turing patterns. Contour pictures of $u_M(x, y)$ plotted for different coupling parameters where $u_M(x,y)$ is defined as the distribution of the maximum values of $u(x, y, t)$ over a time interval $T=20000$ (about 3000 circles of the Rossler oscillations). Localized Turing spots are observed without ambiguity. (a) (b) Spontaneous emergences of two reflection symmetry axes (x- and y- axes) and a symmetry center (0,0). (a) $D_u = D_v = 0.045$, $D_w = 0.25$. (b) $D_u = D_v = 0.047$, $D_w = 0.25$. (c) $D_u = D_v = 0.003$, $D_w = 2.5$. Turing pattern invariant under a discrete translational transformation group $u_M(x+\frac{nL}{4}, y+\frac{mL}{4}) = u_M(x, y), n, m = \pm 1, \pm 2, \pm 3$. (d) $D_u = D_v = 0.024$, $D_w = 2.5$. Turing pattern with a symmetry center while without any mirror symmetry axis. (e) $D_u = D_v = 0.032$, $D_w = 2.5$. Turing pattern with rotational symmetries of $\pi/2$ and $\pi$ rotation angles. (f) $D_u = D_v = 0.03$, $D_w = 2.5$. Turing pattern with discrete mirror reflective, translational and rotational symmetries.

Fig.4

System parameters are the same as Fig. 1(b). (a) Chaotic orbit of an arbitrary space point $(x, y) = (5,5)$. (b) Divergence of two adjacent trajectories with exactly the same initial variables except a $10^{-3}$ deviation at a single site (5, 5). Real line shows the divergence of the two trajectories at site (5, 5), and dished line shows that at site (10, 10) with the largest distance from (5, 5). (c) Function S(x, y) given by Eq.(2). (d) Average absolute difference given by Eq.(3). In (c) and (d) both $S \approx 0$ and $\Delta \approx 0$ represent partial chaos synchronizations.

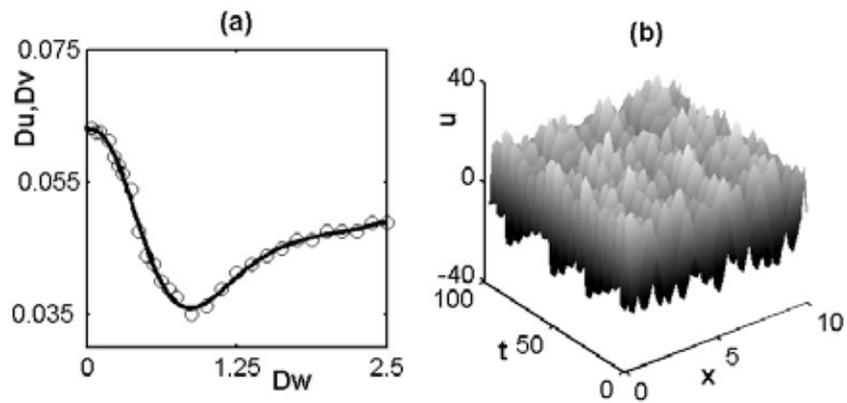

**Fig.1**

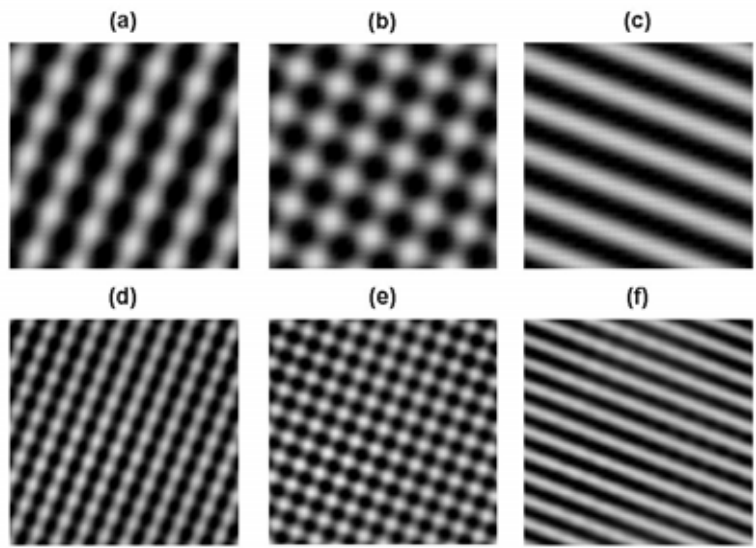

**Fig.2**

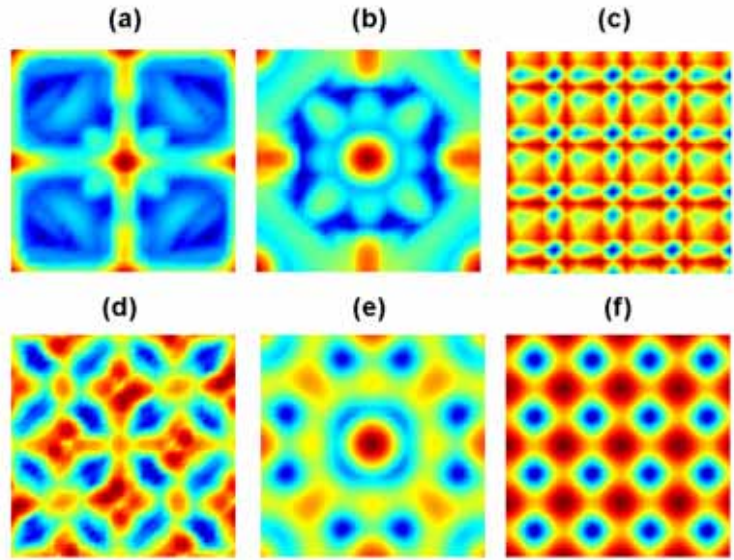

**Fig.3**

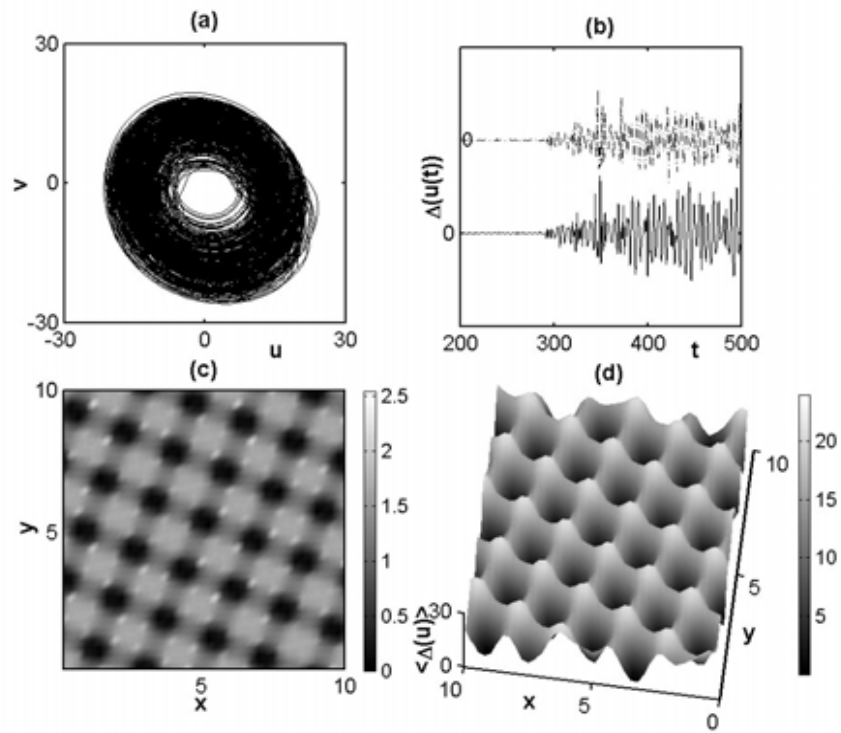

**Fig.4**